\documentclass[preprint,showpacs,preprintnumbers,amsmath,amssymb]{revtex4}

 \usepackage{epsfig}
\usepackage{amssymb}
\usepackage{graphicx}

\begin{document}
\title{Antihydrogen  Gravitational Quantum States }
\author{A. Yu. Voronin, P. Froelich, V.V. Nesvizhevsky}
\affiliation{ P.N. Lebedev Physical Institute, 53 Leninsky
prospect, 117924 Moscow, Russia.
\\
Department of Quantum Chemistry, Uppsala University, Box 518,
SE-75120 Uppsala, Sweden.
\\
Institut Laue-Langevin (ILL), 6 rue Jules Horowitz,
 F-38042, Grenoble, France.}

\begin{abstract}


We present a theoretical study of the motion of  the antihydrogen atom ($\bar{H}$)  in the Earth's gravitational field above a
material surface. We  predict that $\bar{H}$ atom, falling in the Earth's gravitational field  above a material
surface, would settle in long-living quantum states. We point out a method of measuring the difference
in energy of $\bar{H}$ in such states  that allow us to apply  spectroscopy of gravitational levels
based on atom-interferometric principles. We analyze a general feasibility to perform experiments of this kind.
We point out  that such  experiments   provide a method of measuring the gravitational force
($Mg$) acting on $\bar{H}$  and they might be of interest in a context of testing the
Weak Equivalence Principle for antimatter.


\end{abstract}

\maketitle

\section{Introduction}
Galileo, Newton and Einstein  recognized that  all bodies, regardless of their mass and composition,  fall towards the
Earth with an equal gravitational acceleration.
Is that conclusion valid for antimatter? This has never been tested.

In the context of the general relativity theory, the
universality of  free fall is often referred to as the Weak Equivalence
Principle (WEP). Violations of WEP could occur in ordinary matter-matter
interactions e.g. as a result of the difference in the gravitational coupling
to the rest mass and that to the binding energy.   WEP is being tested with increasing sensitivity for macroscopic
bodies. The best test so far confirms WEP
to the accuracy of $2\cdot 10^{-13}$  (using a rotating torsion balance \cite{schl08}).
Ongoing  projects
aim at the accuracy of 1 part in
$10^{16}$  (laser tracking of a pair of test bodies in a  freely falling  rocket \cite{reas10}),  or even to 1 part in
$10^{18}$ (in an Earth orbiting satellite  \cite{over09}). However, in view  of difficulties in unifying the quantum
mechanics with the theory of gravity,  it is of great interest to
investigate  the gravitational properties of {\it quantum mechanical objects},
such as elementary particles or atoms.
Such experiments have been already performed,
e.g. using  interferometric
methods to measure the gravitational acceleration of neutrons \cite{NeutInter1, NeutInter2} and atoms \cite{kase92,pete99,fray04,clad05}.
However, the experiments with {\emph{anti}}atoms ( see \cite{ATRAP,ALPHA} and references therein) are even more interesting in view of testing WEP, because
the theories striving to unify  gravity and quantum mechanics (such as
supersymmetric string theories)
tend  to suggest  violation of the gravitational equivalence of particles
and antiparticles \cite{Sherk}. Experiments testing  gravitational properties of antiatoms   are on the agenda of all experimental
groups working with
antihydrogen (see e.g.  ATHENA-ALPHA \cite{cesa05}, ATRAP \cite{gabr10} and AEGIS
\cite{Aegis1}). One of the challenging aspects in experiments of this kind is to control   the initial parameters of
antiatoms, such as their temperature
and position, with sufficient accuracy \cite{walz04}.

\smallskip

In the present paper we investigate a possibility to explore gravitational properties of antiatoms in the ultimate quantum
limit. We  study antihydrogen atoms levitating in the lowest gravitational states above a material surface.
The existence of such gravitational states for \textit{neutrons} was proven  experimentally
\cite{nesv02,nesv03,nesv05}.
The existence of analogous states for antiatoms seems, at a first glance, impossible because of annihilation of
antiatoms in the  material walls. However, we have  shown   that
ultracold antihydrogen atoms are efficiently reflected from  material surface \cite{voro05l, voro05} due to so-called
quantum reflection from the Casimir-Polder atom-surface interaction potential. We have shown that antihydrogen atoms,
confined  by the quantum
reflection via Casimir forces from below, and  by the gravitational
force from above, would form metastable gravitational quantum states. They  would
bounce on a surface for a finite life-time
(of the order of 0.1 s) \cite{voro05l}.
This simple system can be considered as a
microscopic laboratory for testing the gravitational interaction under extremely
well specified  (in fact, quantized!) conditions.

The annihilation of ultraslow antiatoms in a wall occurs with a small but finite (few percent) probability. It provides a  clear and
easy-to detect signal, which might be used  to measure continuously the antiatom density  in the gravitational states
as a function of time. If antiatoms are settled in a superposition of gravitational states, the    antiatom density
evolves  with beatings, determined by the {\it energy difference} between the gravitational levels. The transition frequencies between the gravitational levels are related to the strength of the gravitational force $Mg$,
acting on antiatoms; here $M$ is the gravitational mass of $\bar{H}$, and $g$ is the Earth's local gravitational field strength.
Also we  show  that a measurement
of {\it differences}  between the energy levels would allow us to disentangle $Mg$ in a way independent on
effects of the  antiatom-surface interaction.

The plan of the paper is the following. In section \ref{gravstates} we study the main properties of the quasi-stationary gravitational states; in section \ref{BounceHbar} we present  the time evolution of the $\bar{H}$ gravitational states superposition; in section \ref{QBE} we discuss a concept of a quantum ballistic experiment, namely  the spatial-temporal evolution of the $\bar{H}$ gravitational states superposition, in section \ref{Feas} we analyze the feasibility of measuring $\bar{H}$ atom properties in gravitational quantum states. In the Appendix we derive useful analytical expressions for the quasi-stationary gravitational  states scalar product.

\section{$\bar{H}$ gravitational states}\label{gravstates}
 In this section we  discuss the properties of the $\bar{H}$ gravitational states   above a material
 surface.

We consider an $\bar{H}$ atom bouncing above a material
surface in the Earth's gravitational field. $\bar{H}$ is confined
 due to the quantum scattering from the  Casimir-Polder
potential  below, and  the gravitational field  above.
The Schr\"{o}dinger equation for the $\bar{H}$ wave-function $\Psi(z)$ in such a superposition of atom-surface and
gravitational potentials is:
\begin{equation} \label{Schr}
\left[ -\frac{\hbar^2 \partial ^{2}}{2m\partial z^{2}}+V(z)+Mgz-E\right] \Psi
(z)=0
\end{equation}
Here $z$ is the distance between the surface and the $\bar{H}$ atom, and $V(z)$ is the atom-surface interaction
potential with a long-range asymptotic form $V(z)\sim-C_4/z^4$.  We distinguish between the gravitational mass, that
  we refer to as $M$ and the inertial mass,  denoted by
  $m$ hereafter. The wave-function $\Psi(z)$ satisfies the full absorption boundary condition at the surface ($z=0$)
  \cite{voro05}, which stands for the annihilation of antiatoms in the material wall.

The
characteristic length and energy scales are
\begin{eqnarray}\label{scaleL}
l_0 &=&\sqrt[3]{\frac{\hbar^{2}}{2mMg}},\\
l_{CP}&=&\sqrt{2mC_4}, \label{scaleCP}\\
\varepsilon_0 &=&\sqrt[3]{\frac{\hbar^2M^2g^2}{2m}},\label{scaleE}\\
\varepsilon_{CP}&=&\frac{\hbar^2}{4 m^2 C_4}.\label{scaleEcp}
\end{eqnarray}

%
 Here $l_0 =5.871$ $\mu m$ is the characteristic gravitational length scale, $l_{CP}=0.027$ $\mu m$ is the
 characteristic Casimir-Polder interaction length scale, $\varepsilon_0=2.211$ $10^{-14}$
  a.u. is the characteristic  gravitational energy scale, and $\varepsilon_{CP}=1.007$ $10^{-9}$ a.u. is   the
  Casimir-Polder energy scale. As one can see, the gravitational length scale is much larger than the Casimir-Polder
  length scale $l_0\gg l_{CP}$, while the gravitational energy scale is much smaller than the Casimir-Polder energy
  scale $\varepsilon_0\ll \varepsilon_{CP}$. It is useful to introduce the gravitational time scale $\tau_0$:
\begin{equation}\label{tau0}
\tau_0=\hbar/\varepsilon_0\simeq 0.001s
\end{equation}

For large atom-surface separation distances $z\gg l_{CP}$ the solution of eq.(\ref{Schr}) has a form:
\begin{equation}\label{gravfree}
\Psi(z)\sim \mathop{\rm Ai}( \frac{z}{l_0}-\frac{E}{\varepsilon_0})+K(E)\mathop{\rm
Bi}(\frac{z}{l_0}-\frac{E}{\varepsilon_0})
\end{equation}
where $\mathop{\rm Ai}(x)$ and $\mathop{\rm Bi}(x)$ are the Airy functions \cite{abra72}.
The requirement of square integrability of the wave-function  $\Psi(z\rightarrow \infty)\rightarrow 0$ results in
the following equation for the energy levels of the gravitational states in the presence of the Casimir-Polder
interaction:
\begin{equation}\label{eigenexact}
K(E_n)=0
\end{equation}

 The hierarchy of the Casimir-Polder  and gravitational scales $l_{CP}\ll l_0$ suggests that  the quantum reflection
 from the Casimir-Polder potential can be accounted for by modifying  the boundary condition for the quantum bouncer (a
 particle  bouncing in the gravitational field above a surface, the interaction of the latter with a particle is modeled
 by infinite reflecting wall). The quantum bouncer wave-function satisfies the following equation system:

 \begin{equation}\label{gravEq}
 \left\{\begin{array}{cll}\left[ -\frac{\hbar^2 \partial ^{2}}{2m\partial z^{2}}+Mgz-E_n\right] \Phi_n
(z)=0\\
\Phi_n(z= 0)= 0
\end{array}
\right.
\end{equation}
The quantum bouncer energy levels  are known to be equal to \cite{nesv02}:
\begin{eqnarray}\label{En0}
E_n^0&=&\varepsilon_0 \lambda_n^0, \\ \label{lambda0}
 \mathop{\rm Ai}(-\lambda_n^0)&=&0.
\end{eqnarray}

 Table \ref{Table1} summarizes the eigenvalues and classical turning points $z_n^0=E_n^0/(Mg)$ for the first seven
 gravitational states of a quantum bouncer (with the mass of antihydrogen).

 \begin{table}
 \centering
 \begin{tabular}{|c|l|l|l|}
  \hline
  $n$ & $\lambda_n^0$ &  $E_n^0$, peV & $z_n^0$, $\mu m$\\
  \hline
  1 & 2.338 &  1.407 & 13.726 \\
  2 & 4.088 &  2.461 & 24.001\\
  3 & 5.521 &  3.324 & 32.414\\
  4 & 6.787 &  4.086 & 39.846\\
  5 & 7.944 &  4.782  & 46.639 \\
  6 & 9.023 &  5.431  & 52.974\\
  7 & 10.040&  6.044  & 58.945 \\
  \hline
 \end{tabular}
 \caption{The eigenvalues, gravitational energies and classical turning
 points of a quantum bouncer with a mass of (anti)hydrogen in the Earth's gravitational field.
 } \label{Table1}
 \end{table}


 For the distances $l_{CP}\ll z \ll l_0$ one could  neglect the gravitational potential in eq.(\ref{Schr}). In this
 approximation, the solution of eq.(\ref{Schr}) has  the following asymptotic form:
 \begin{equation}\label{free}
 \Psi(z)\sim \sin(k z +\delta(E)).
 \end{equation}
 Here $k$ is the wave vector $k=\sqrt{2mE}$, and $\delta(E)$ is the phase-shift  of $\bar{H}$  reflected from the
 Casimir-Polder potential \emph{in  absence of the gravitational field} \cite{voro05}. Matching  asymptotics in
 eq.(\ref{free}) and eq.(\ref{gravfree}) we get a
 relation between the phase-shift  $\delta(E)$ and the $K-$function introduced in Eq.(\ref{gravfree}):
 \begin{equation}\label{Kapprox}
 K(E)=-\frac{\tan(\delta(E))\mathop{\rm Ai'}(-E/\varepsilon_0)-k l_0 \mathop{\rm
 Ai}(-E/\varepsilon_0)}{\tan(\delta(E))\mathop{\rm Bi'}(-E/\varepsilon_0)-k l_0 \mathop{\rm Bi}(-E/\varepsilon_0)}.
 \end{equation}
In deriving the above expression we took into account that the relation between  $K(E)$ and $\delta(E)$ should not depend on the matching point $z_m$  and thus can be formally attributed to $z_m=0$.
An equation  for the distorted gravitational levels could be obtained by substitution of eq.(\ref{Kapprox}) into
eq.(\ref{eigenexact}):
 \begin{equation}\label{BC}
\frac{\tan(\delta(E_n))}{k l_0}=\frac{\mathop{\rm Ai}(-E_n/\varepsilon_0)}{\mathop{\rm Ai'}(-E_n/\varepsilon_0)}.
\end{equation}
This equation is equivalent to the following boundary condition:
\begin{equation}\label{BCmodified}
\frac{\Phi( 0)}{\Phi'( 0)}=\frac{\tan(\delta(E_n))}{k }.
\end{equation}
Thus the following equation system  describes an $\bar{H}$ atom, confined by the Earth's gravitational field and the quantum
reflection from the Casimir-Polder potential:
 \begin{equation}\label{gravEqmod}
 \left\{\begin{array}{cll}\left[ -\frac{\hbar^2 \partial ^{2}}{2m\partial z^{2}}+Mgz-E_n\right] \Phi_n
(z)=0\\
\frac{\Phi( 0)}{\Phi'( 0)}=\frac{\tan(\delta(E_n))}{k }
\end{array}
\right.
\end{equation}

For the lowest gravitational states the condition $k l_{CP} \ll 1$ is valid. Thus the scattering length approximation
for the phase-shift $\delta(E)\approx -k a_{CP}$ is well justified. The \emph{complex-value} quantity \cite{voro05}:
\begin{eqnarray}
a_{CP}&=&-(0.10+i1.05)l_{CP},\\
a_{CP}&=&-0.0027-i0.0287 \mu m
\end{eqnarray}
 is the scattering length on the Casimir-Polder potential provided full absorbtion in the material wall.

Thus the equation for the lowest eigenvalues (\ref{BC})  has a form:
\begin{equation}\label{BClow}
\frac{a_{CP}}{l_0}=-\frac{\mathop{\rm Ai}(-E_n/\varepsilon_0)}{\mathop{\rm Ai'}(-E_n/\varepsilon_0)}.
\end{equation}

The above equation is equivalent to the following boundary condition for the wave-function $\Phi(z)$ of a particle in
the gravitational potential eq.(\ref{Schr}):
\begin{equation}\label{BClow1}
\Phi(z\rightarrow 0)\rightarrow z-a_{CP}
\end{equation}

Because of the imaginary part of the scattering length $a_{CP}$, the gravitational states of $\bar{H}$ above a material
surface are \emph{quasi-stationary decaying} states.
For low quantum numbers $n$, it is easy to relate the  lowest quasi-stationary energy levels $E_n$ to the unperturbed gravitational energy levels $E_n^0$ of a quantum bouncer. Indeed the  variable substitution $z=\widetilde{z}+a_{CP}$ transforms
Eq.(\ref{gravEq},\ref{BClow1}) to the equation system for the quantum bouncer:
 \begin{eqnarray}\label{gravEq1}
 \left[ -\frac{\hbar^2\partial ^{2}}{2m\partial \widetilde{z}^{2}}+Mg\widetilde{z}-(E_n-Mg a_{CP})\right] \Phi_n
(\widetilde{z})=0\\
\Phi_n(\widetilde{z}\rightarrow 0)\rightarrow 0 \label{BC1}
\end{eqnarray}

The eigenvalues $E_n$ and eigenfunctions $\Phi_n$ are :
\begin{eqnarray}\label{energy}
E_n=E_n^0+Mga_{CP},\\
\Phi_n(z)=\frac{1}{N_i}\mathop{\rm Ai}((z-a_{CP})/l_0-\lambda_n^0),\label{Phi}
\end{eqnarray}
where $N_i$ is the normalization coefficient (see Eqs.(\ref{normex},\ref{norm}) in the Appendix).
In the following, we will use the dimensionless eigenvalues $\lambda_n=E_n/\varepsilon_0$:

\begin{equation}\label{lambdan}
\lambda_n=\lambda_n^0+a_{CP}/l_0
\end{equation}

 An important message from  the above expression is that the  complex shift $ Mg a_{CP}$
(due to the  account of  quantum reflection on the Casimir-Polder potential) is \emph{the same} for all  low-lying  quasi-stationary gravitational levels.   It means that  the transition frequencies between the gravitational states are not affected by the Casimir-Polder interaction, provided  the latter can be described by the complex scattering length $a_{CP}$. The scattering  approximation is valid in the limit $k_n a_{CP}\rightarrow 0$, where $k_n=\sqrt{2mE_n}$ (let us note that for the first gravitational state $|k_1 a_{CP}|=0.0071$). However, accounting for the higher order $k$-dependent terms in  Eq.(\ref{BC}) would result in the state dependent shift of the gravitational states due to the Casimir-Polder interaction. We use a known low energy expansion of the $s$-wave phase-shift $\delta(E)$ in a homogeneous $1/z^4$ potential \cite{EffRad}, in which we keep the two leading $k$-dependent terms:
\begin{equation}
\tilde{a}_{CP}(k) \cot\left(\delta(k)\right)\simeq
-1+\frac{\pi}{3}\frac{l_{CP}}{a_{CP}}(l_{CP}k)+\frac{4}{3}(l_{CP}k)^2\ln\left(\frac{l_{CP}k}{4}\right)+...
\end{equation}
We introduce a k-dependent modified "scattering length" $\tilde{a}_{CP}(k)\equiv-\delta(k)/k$ and get the following
expression for $\tilde{a}_{CP}(k)$:

\begin{equation}\label{ak}
\tilde{a}_{CP}(k)\simeq a_{CP}+\frac{\pi}{3}l_{CP} (l_{CP}k)+\frac{4}{3}a_{CP}(l_{CP}k)^2\ln\frac{l_{CP}k}{4}
\end{equation}
The  leading k-dependent term in the above expression  $\frac{\pi}{3}l_{CP} (l_{CP}k)$ is real and independent on
properties of the inner part of the Casimir-Polder interaction. It is determined by the asymptotic form of the
potential, thus it depends on the Casimir-Polder length scale $l_{CP}$ only.
Then the modified equation for the gravitational state energies is:
\begin{equation}
E_n=E_n^0+Mg \tilde{a}_{CP}(E_n)
\end{equation}
Taking into account the smallness of the k-dependent terms (for the lowest gravitational states) in expression (\ref{ak}),
we get:
\begin{equation}
E_n\simeq E_n^0+Mg \tilde{a}_{CP}(k_n^0)= \varepsilon\left(\lambda_n^0+ a_{CP}/l_0+\frac{\pi l_{CP}}{3 l_0}
(l_{CP}k_n^0)+\frac{4 a_{CP}}{3l_0}(l_{CP}k_n^0)^2\ln\frac{l_{CP}k_n^0}{4}\right).
\end{equation}
Here $k_n^0=\sqrt{2 m E_n^0}$.

The account for k-dependent terms in Eq.(\ref{ak})  modifies the transition frequencies between the gravitational
states in a way, dependent on the Casimir-Polder interaction. However, such modification is  very weak. Indeed, taking
into account, that $l_{CP} k_n^0\sim l_{CP}/l_0$ for the lowest gravitational states , the leading k-dependent term
corrections to the gravitational energy are of the second order in a small parameter $l_{CP}/l_0$. The transition
frequency  between the first and  second gravitational states equals  $\omega_{12}=\omega_{12}^0+\Delta_{12}$, where
$\omega_{12}^0=(E_2^0-E_1^0)/(2\pi\hbar)=254.54$ Hz, and $\Delta_{12}=Mg(\tilde{a}_{CP}(k_2^0)-\tilde{a}_{CP}(k_1^0))=
0.0017$ Hz.

 An account of  the  first two  terms in Eq.(\ref{ak}) provides equal decay width for the lowest gravitational states. This
 width is determined by the  probability of antihydrogen penetrating to the surface and annihilating.

\begin{equation}
\Gamma_n= \varepsilon \frac{b}{2l_0}.\label{Wgrav}
\end{equation}
Here we use a standard notation $b=4\mathop{\rm Im}a_{CP}$:
\begin{equation} b=0.115 \mbox{ } \mu m.
\end{equation}
The widths of the gravitational states
(\ref{Wgrav}) are proportional to the ratio
$\varepsilon/l_0$.  Using Eqs. (\ref{scaleL}) and (\ref{scaleE}) we could find
that this ratio is equal to the gravitational force
$\varepsilon/l_0=Mg$ so that

\begin{equation}
\Gamma_n = \frac{b}{2} Mg.
\end{equation}

 The corresponding life-time (calculated for an ideal conducting surface) is

\begin{equation}
\tau=\frac{2\hbar}{Mgb} \simeq 0.1 \mbox{ s}. \label{time}
\end{equation}
We note factorization of the gravitational effect (appearing in the
above formula via  a factor $Mg$) and  the quantum
reflection effect, manifestating through the constant $b$. Such a factorization
is a consequence of the smallness of the ratio of the characteristic scales
$b/(2l_0)\simeq 0.01$.

Comparing the $\bar{H}$ lifetime in the lowest gravitational states  with the classical period $T=2\sqrt{\frac{2l_0
\lambda_1}{g}}\simeq 0.0033$ s of  $\bar{H}$  with the ground state energy bouncing in the gravitational field, we see
that $\bar{H}$ bounces in average about $30$ times before  annihilating. This  shows that the lowest
gravitational states  are well resolved quasi-stationary states.

\begin{figure}
 \centering
\includegraphics[width=120mm]{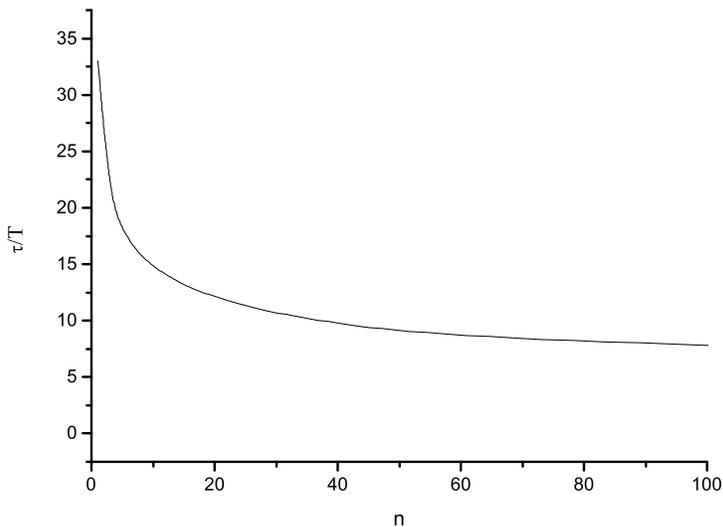}
\caption{The number of $\bar{H}$ bounces during the lifetime of $n-$th gravitational state}\label{resol}
\end{figure}

It is interesting to estimate the maximum gravitational quantum number $N$, below which the gravitational states are
still  resolved, i.e:
\begin{equation}\label{resolve}
\frac{\tau_{N}}{T_{N}}=\frac{2\pi \hbar}{\Gamma(N)\frac{dE(n)}{dn}}> 1.
\end{equation}
Here $\tau_{N}$ is the lifetime of the $N-$th gravitational state, $T_N$ is a classical period, corresponding to the
$N-$th state via $T_N=2\hbar\pi/(dE(n)/dn)$.
For such an estimation we transform eq.(\ref{BC}) using the  asymptotic form of the Airy function  for a large negative
argument and get:
\begin{equation}\label{Esemiclass}
\lambda_n=\left( \frac{3}{2}(\pi [n-\frac{1}{4}]-\delta(E_n))\right)^{2/3}.
\end{equation}
The accuracy of the above equation increases with increasing $n$;  it gives the energy value within a few percent even
for $n=1$. In the energy domain of interest $|\delta(E)|\ll \pi [n-\frac{1}{4}]$ \cite{voro05}, so:
\begin{equation}\label{Esemiclass1}
\lambda_n\simeq \lambda_n^0-\frac{\delta(E_n^0)}{\sqrt{\lambda_n^0}}.
\end{equation}
  Here we used the semiclassical approximation for $\lambda_n^0$ \cite{NVP}:
  \begin{equation}
  \lambda_n^0\simeq \left( \frac{3}{2}\pi [n-\frac{1}{4}]\right)^{2/3}
  \end{equation}
One could verify that in the case of small $n$ the above equation reduces to eq.(\ref{lambdan}). Substitution of
eq.(\ref{Esemiclass1}) into eq.(\ref{resolve}) results to:
\begin{equation}\label{resolve1}
\frac{\tau_{n}}{T_{n}}\simeq \frac{1}{4\mathop{\rm Im}\delta(E_n^0)}.
\end{equation}
The ratio $\tau(n)/T(n)$ expresses the number of $\bar{H}$ classical bounces during the lifetime of the $n$-th
state. This dependence  is shown in Fig.\ref{resol}.
Using numerical values $\delta(E)$, calculated in \cite{voro05}, we find that inequality (\ref{resolve}) holds for
\begin{equation}
n<N=30000.
\end{equation}
The corresponding energy $E_{N}=6$ $10^{-11}$ a.u., and the characteristic size of such states is as large as
$H_{N}=1.6$ $cm$.
This means that the concept of the quasi-bound gravitational states is justified  not only for the lowest states, but it
also might be applied for highly excited states.

The quasi-stationary character of the antiatom gravitational states above a material surface manifests itself in a
nonzero current through the bottom  surface ($z=0$). Indeed, the expression for the current is:
\begin{equation}\label{j}
j(z,t)=\frac{i\hbar}{2M}\left(\Phi(z,t) \frac{d \Phi^*(z,t)}{dz}-\Phi^*(z,t) \frac{d \Phi(z,t)}{dz}\right).
\end{equation}
taken at $z=0$ for a given gravitational state (\ref{Phi}) turns out to be equal to:
\begin{equation}\label{j0}
j(0,t)=\varepsilon \exp(-\frac{\Gamma}{\hbar} t) \frac{\mathop{\rm Ai^*}(-\lambda_n)\mathop{\rm
Ai'}(-\lambda_n)-\mathop{\rm Ai}(-\lambda_n) \mathop{\rm Ai'^*}(-\lambda_n) }{N_iN_i^*}.
\end{equation}
Here $\lambda_n$ is given by Eq. (\ref{lambdan}), $N_i$ is the normalization factor. We take into account the smallness of the ratio $a_{CP}/l_0$ and
Eq.(\ref{lambda0}), and get :

 \begin{equation}
 \mathop{\rm Ai}(-\lambda_n)\approx-\frac{a_{CP}}{l_0}\mathop{\rm Ai'}(-\lambda_n^0),
 \end{equation}
 which is exact up to the second order in the ratio $a_{CP}/l_0$. Now taking into account an explicit form of the
 normalization coefficients (Eq.(\ref{norm}) in Appendix) $N_i=\mathop{\rm Ai'}((-\lambda_n^0)$, we  get finally:
 \begin{equation}\label{jex}
 j(0,t)=-\varepsilon \frac{b}{2\hbar l_0}\exp(-\frac{\Gamma}{\hbar} t)=-\frac{\Gamma}{\hbar} \exp(-\frac{\Gamma}{\hbar}
 t).
 \end{equation}
This result is in full agreement with Eq.(\ref{Wgrav}) as far as:
\begin{equation}
\frac{d}{dt}\int_0^{\infty}|\Phi(z,t)|^2dz=j(0,t)=-\frac{\Gamma}{\hbar}\exp(-\frac{\Gamma}{\hbar} t).
\end{equation}

\section{Bouncing antihydrogen}\label{BounceHbar}
In this section, we are interested in the evolution of an initially prepared  arbitrary superposition of several lowest
gravitational states of $\bar{H}$. In the following, we will limit our treatment to the scattering length
approximation for describing the Casimir-Polder interaction, and will neglect all, except the first, term in the
expression (\ref{ak}), so that $\tilde{a}_{CP}(k)\approx a_{CP}$. The corresponding $\bar{H}$ wave-function is:
\begin{equation}\label{superpos}
\Phi(z,t)=\sum_{i=1}^n \frac{C_i}{N_i}\mathop{\rm Ai}(z/l_0-\lambda_i)\exp(-i  \lambda_i \frac{t}{\tau_0}).
\end{equation}
Here $\tau_0$ is the characteristic $\bar{H}$ bouncer time scale, $C_i$ are expansion coefficients and
$N_i=\mathop{\rm Ai'}(-\lambda_i)$ are the normalization factors of the gravitational states (see the Appendix).

We are interested in the evolution of the number of antihydrogen atoms  as a
function of time:
\begin{equation}\label{Flux}
F(t)=\int_0^\infty |\Phi(z,t)|^2dz= \sum_{i,j=1}^n \int_0^\infty \frac{C_{j}^*C_i}{N_j^*N_i}\mathop{\rm
Ai^*}(z/l_0-\lambda_{j})\mathop{\rm Ai}(z/l_0-\lambda_i)\exp(-i\varepsilon (\lambda_i-\lambda_{j}^*)t)dz
\end{equation}

First, let us  note that  the above expression for the total number of particles is no longer constant because of the decay
of the quasi-stationary gravitation states.
Second, the quasi-stationary   gravitational states corresponding to  different energies are  non-orthogonal:
\begin{equation}
\frac{1}{N_iN_j}\int_0^\infty\mathop{\rm Ai^*}(z/l_0-\lambda_{j})\mathop{\rm Ai}(z/l_0-\lambda_i) dz\equiv
\alpha_{ij}\neq\delta_{ij}.
\end{equation}
In the Appendix we will derive the following expression for the cross-terms $\alpha_{ij}$, exact up to the second order of
the small ratio $a_{CP}/l_0$:
\begin{equation}\label{crossTA}
\alpha_{i\neq j}=i\frac{b/(2l_0)}{\lambda_j^0-\lambda_i^0+i b/(2l_0)}
\end{equation}
As one can see, such cross-terms vanish if there is no decay, i.e. if $b=4\mathop{\rm Im}a_{CP}\rightarrow 0$.

Now we can calculate an expression for the number of antihydrogen atoms as a function of time (\ref{Flux}):
\begin{equation}\label{Ft}
F(t)=\exp(-\frac{\Gamma}{\hbar} t) \left(\sum_{i}^n |C_i|^2+ 2\mathop{\rm Re}\sum_{i>j}^n\sum_{j}^n C_{j}^*C_{i}\frac{i
b/(2l_0)}{\lambda_j^0-\lambda_i^0+i b/(2l_0)} \exp(-i(\lambda_i^0-\lambda_j^0)\frac{t}{\tau_0})\right).
\end{equation}

 From Eqs.(\ref{Ft}) and (\ref{Wgrav}) we get the following expression for the disappearance (annihilation) rate
 $-\frac{dF(t)}{dt}$, keeping the terms up to the second order in the ratio $a_{CP}/l_0$:
 \begin{equation}\label{Nt}
 \frac{dF(t)}{dt}=-\frac{\Gamma}{\hbar} \exp(-\frac{\Gamma}{\hbar} t) \left(\sum_{i}^n |C_i|^2+2\mathop{\rm
 Re}\sum_{i>j}^n\sum_{j}^n C_{j}^*C_{i}\exp(-i(\lambda_i^0-\lambda_j^0)\frac{t}{\tau_0})\right).
 \end{equation}

 For a superposition of the two gravitational states  with the equal coefficients $C_{1,2}$ (say $C_1=C_2=1$), the
 above expression gets a simple form:
 \begin{equation}\label{2st}
 \frac{dF_{12}(t)}{dt}=-\frac{\Gamma}{\hbar} \exp(-\frac{\Gamma}{\hbar} t)\left (1+\cos(\omega_{12}t)\right ).
 \end{equation}
Here $\omega_{12}=(\lambda_2^0-\lambda_1^0)/\tau_0$.
The same result could be obtained by calculating the flux $j(0,t)$ Eq.(\ref{j}) for a superposition of states
(\ref{superpos}).

\begin{figure}
 \centering
\includegraphics[width=100mm]{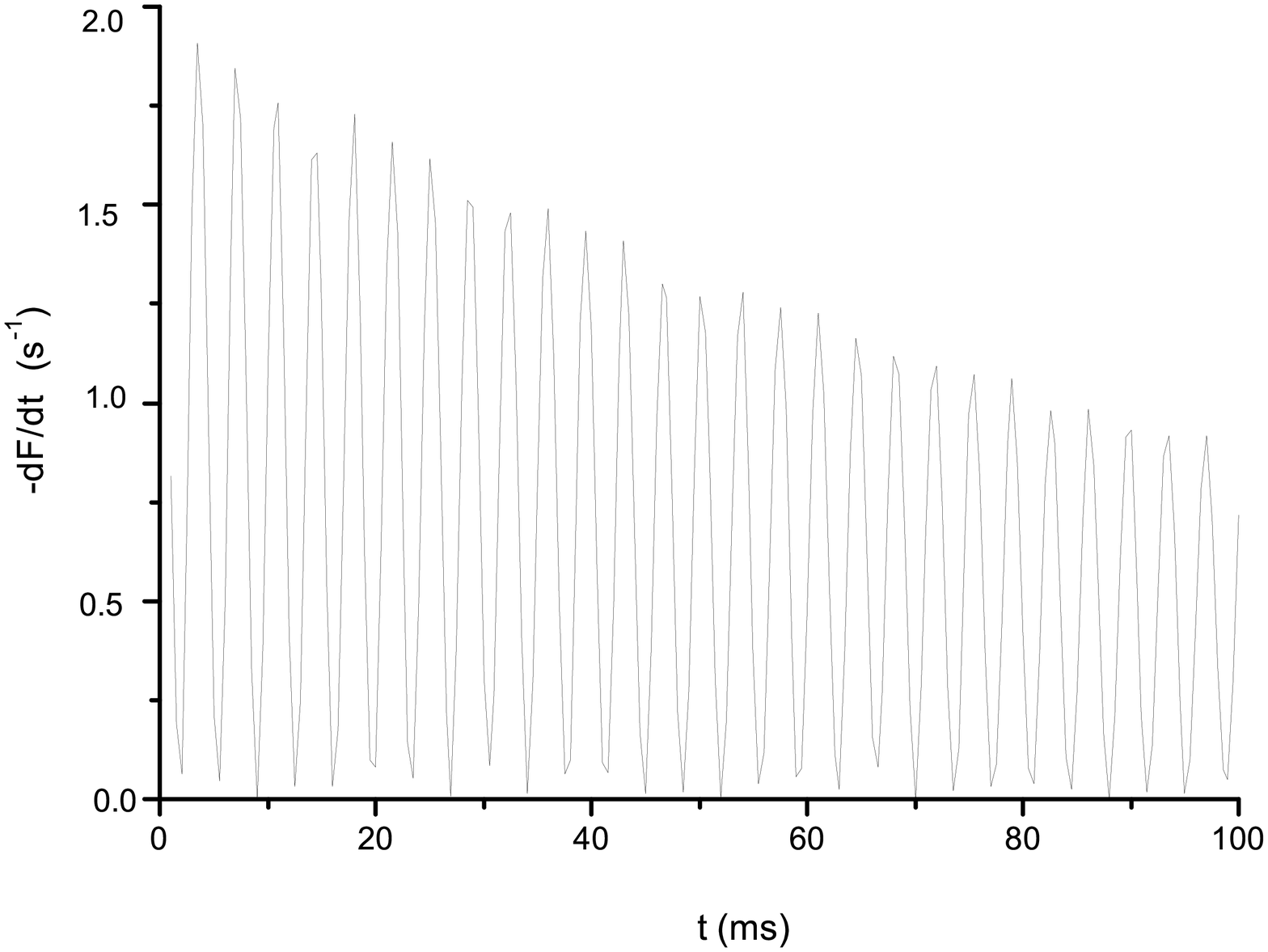}
\caption{Evolution of the annihilation rate of $\bar{H}$ atom    in a superposition of the first and the  second
gravitational states.} \label{twost}
\end{figure}

 One can see, that the disappearance rate decays as a function of time according to the exponential law with the width
 $\Gamma$ (the same for the lowest states), also it oscillates   with the transition frequency between the first and
 second gravitational states (equal to $254.54$ Hz). We plot in Fig.\ref{twost} the time evolution of $\bar{H}$
 disappearance rate in a superposition of two lowest states. Curiously, the oscillation of disappearance rate is  the direct consequence of decaying character of gravitational states. Indeed, such an oscillation is observable due to the nonvanishing  contribution of the interference term in the expression for the total probability to find antihydrogen atoms, given by Eq.(\ref{Ft}). As one can see from Eq.(\ref{crossTA}) this contribution is proportional to the imaginary part of the scattering length, it would vanish in case there were no decay of gravitational states due to annihilation in the material wall, described by parameter $b/l_0$.
\begin{figure}
 \centering
\includegraphics[width=100mm]{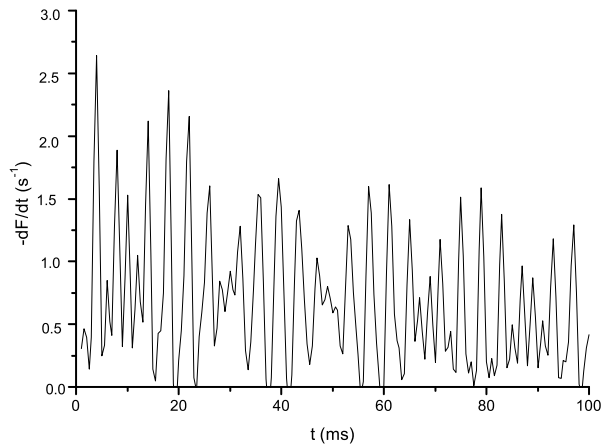}
\caption{Evolution of the annihilation rate of $\bar{H}$ atom    in a superposition of first, second and third
gravitational states.} \label{threest}
\end{figure}

 So far the oscillation frequency of the disappearance rate $N(t)$ corresponds to the energy difference between the
 unperturbed gravitational levels. Expression (\ref{Nt}) does not include the shift of gravitational state energies
 $\mathop{\rm Re}a_{CP}/l_0$ as it is equal for all the gravitational states, thus it is canceled out in the energy
 difference. The account for higher order k-dependent terms in (\ref{ak}) would result in a small (second order of the
 ratio $(a_{CP}/l_0)$) correction to the transition frequency.
A measurement of  the oscillation frequency $\omega_{12}$ given by Eq. (\ref{2st}) would allow us to extract  the
following combination of the gravitational and the inertial masses from eq.(\ref{scaleE}):
\begin{equation}\label{Mm}
\frac{M^2}{m}=\frac{2 \hbar \omega_{12}^3}{g^2(\lambda_2^0-\lambda_1^0)^3}.
\end{equation}
 Under the additional assumption of the equality of the known inertial mass of the
 \emph{hydrogen} atom $m_H$ and that of antihydrogen, imposed by CPT, we get:
 \begin{equation}
 M=\sqrt{\frac{2 m_H \hbar \omega_{12}^3}{g^2(\lambda_2^0-\lambda_1^0)^3}}.
 \end{equation}

 The  evolution of \emph{three} gravitational state superposition provides information not only about the characteristic
 energy scale $\varepsilon_0$ but also about the  level spacing as a function of quantum number $n$, characterized by
 the value $d^2E(n)/dn^2$. Such a study might be interesting for testing additional (to Newtonian gravitation) interactions (see \cite{Axion,
 ShortRange}  and references there in) between $\bar{H}$ and a material surface with the characteristic spatial scale
 of the order of micrometers. Such interactions would manifest as nonlinear additions to the gravitational potential,
 which would modify the spectrum character. In the case of three state superposition, the disappearance rate (\ref{Nt})
 has the form:
 \begin{equation}\label{3st}
 \frac{dF_{123}(t)}{dt}=-\frac{2}{3}\frac{\Gamma}{\hbar} \exp(-\frac{\Gamma}{\hbar}
 t)\left(\frac{3}{2}+\cos(\omega_{12}t)
 +\cos(\omega_{23}t)+\cos((\omega_{12}+\omega_{23})t)\right).
 \end{equation}
Here $\omega_{ij}=(\lambda_j^0-\lambda_i^0)/\tau_0$.
One could verify that the period of coherence of $\cos(\omega_{12}t)$ and $\cos(\omega_{23}t)$ terms is:
\begin{equation}\label{Trev}
T_{r}=\frac{2\pi}{\omega_{12}-\omega_{23}}\simeq 0.02 s.
\end{equation}
A semiclassical expression for $T_r$ is:
  \begin{equation}
  T_r\approx \frac{2\pi}{|d^2E/dn^2|}.
  \end{equation}
One can see that the period $T_r$ is a quantum limit analog of a half revival period $T_{rev}=4\pi/|d^2E/dn^2|$
($T_{rev}$ characterizes the time period after which the evolution of the wave-packet returns to the semiclassical
behavior, see \cite{Reviv} for details  and reference therein).  In Fig.\ref{threest} we plot the annihilation events
as a function of time (\ref{3st}) for a superposition of three lowest gravitational states. The period $T_r$ is clearly
seen as a period of modulation of a rapidly oscillating function. The ratio
\begin{equation}
T_r/\tau_0=\frac{2\pi}{\lambda_3-2\lambda_2+\lambda_1}
\end{equation}
is sensitive to  any nonlinear addition to the  gravitational potential. Indeed while linear corrections to gravitational potential can only change $\varepsilon_0$, nonlinear additions change the derivative of levels density $|d^2E/dn^2|$.

\section{Quantum ballistic experiment}\label{QBE}
Two independent experiments are needed in order to determine  the gravitational mass $M$ and the inertial mass $m$ of antihydrogen. In the
previous section, we showed that a combination of gravitational and inertial masses $M$ and $m$, given by Eq.(\ref{Mm}),
can be extracted from the frequency measurement eq.(\ref{2st}, \ref{3st}). An independent information could be obtained
from measurement of the spatial density distribution of $\bar{H}$ in a superposition of the gravitation states, for
instance, in the flow-throw experiment (a kind of a beam scattering experiment), in which  $\bar{H}$-atoms with a wide horizontal velocity  distribution move along
the mirror surface. The time of flight along the mirror should be
measured simultaneously with the spatial density distribution in a position-sensitive detector, placed at the mirror
exit. Such a detector would be able to measure the density distribution along the vertical axis at a given time instant. The horizontal component of $\bar{H}$ motion could be treated classically. Due to a broad distribution of
horizontal velocities in the beam, atoms would be detected within a wide range of time intervals between
their entrance to the mirror and their detection at the exit. In such an approach, we could study the time evolution of
$\bar{H}$ probability density at a given position $z$:
\begin{eqnarray}\label{Pzt}
|\Phi_{(12)}(z,t)|^2&=&\exp(-\frac{\Gamma}{\hbar} t)\left( |\Phi_{(12)}^{av}(z)|^2
+2\mathop{\rm Re}\Phi_{(12)}^{int}(z)\exp(-i\omega_{12} t)\right) \\
|\Phi_{(12)}^{av}(z)|^2&=&\left|\frac{\mathop{\rm Ai}(z/l_0-\lambda_1)}{\mathop{\rm Ai'}(-\lambda_1)}\right|^2+
\left|\frac{\mathop{\rm Ai}(z/l_0-\lambda_2)}{\mathop{\rm Ai'}(-\lambda_2)}\right|^2 \\
\Phi_{(12)}^{int}(z)&=& \frac{\mathop{\rm Ai}(z/l_0-\lambda_1)\mathop{\rm Ai}(z/l_0-\lambda_2)}{\mathop{\rm Ai'}(-\lambda_1)\mathop{\rm Ai'}(-\lambda_2)}
\end{eqnarray}

\begin{figure}
 \centering
\includegraphics[width=100mm]{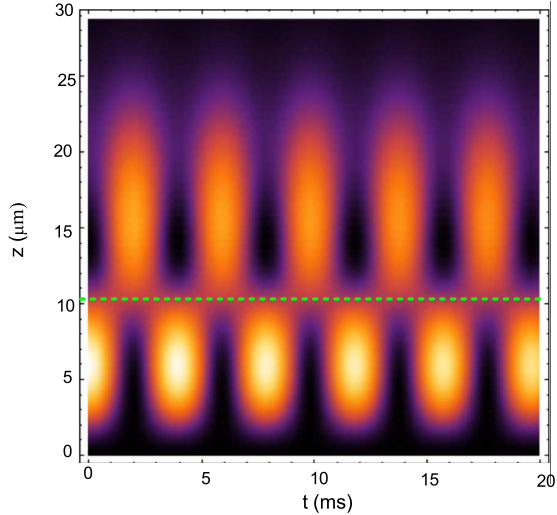}
\caption{Color.The  probability density  of $\bar{H}$ in a superposition of the first and second gravitational states,
as a function of the height $z$ above a mirror (vertical-axis) and the time $t$ (horizontal-axis).
Dark shade: low probability density. Light shade: high probability density.
The dashed line indicates the position of the node in the wave-function of the second state} \label{Fig2zt}
\end{figure}

The transition $\omega_{12}=254.54 Hz$ could be extracted from the probability density time evolution at a given $z$.
The length scale $l_0$ could be extracted from the position of the zero $z_1^{(2)}$ of the wave-function. Here
superscript stands for the state number, and a subscript corresponds to the number of zero for a given state. Thus such
a position is determined by the condition  $\mathop{\rm Ai}(z_1^{(2)}/l_0-\lambda_2)=0$; is equal to the following
expression:
 \begin{equation}
 z_1^{(2)}=(\lambda_2-\lambda_1)l_0=10.27\mu m.
 \end{equation}
The probability density in Eq.(\ref{Pzt}) of a two states superposition at $z=z_1^{(2)}$ behaves like the probability
density of the ground state alone:
\begin{equation}
|\Phi_{12}(z_1^{(2)},t)|^2=\exp(-\frac{\Gamma}{\hbar} t)\left|\frac{\mathop{\rm
Ai}(z_1^{(2)}/l_0-\lambda_1)}{\mathop{\rm Ai'}(-\lambda_1)}\right|^2.
\end{equation}
The probability density at a height $z_1^{(2)}$ does not exhibit any time-dependent oscillations. We show the
probability density  as a function of the height $z$ above a mirror (y-axis) and the time $t$ (x-axis) in a
superposition of the first and second gravitational states in Fig.\ref{Fig2zt}.

The position of the node in the wave-function of the second state is shown in Fig.\ref{Fig2zt} as a horizontal line, separating lower and upper rows of periodic maxima and minima in the probability
density plot. The position $z_1^{(2)}$ does not depend on initial populations of the gravitational states, which makes it
beneficial for extracting the spatial scale $l_0$.

From knowing the length $l_0$ and the energy $\varepsilon_0$ scales, one could get the following expressions for the
inertial $m$ and gravitational $M$ masses of $\bar{H}$:
\begin{eqnarray} \label{mM}
m&=&\frac{\hbar^2}{2\varepsilon_0 l_0^2},\\
M&=&\frac{\varepsilon_0}{gl_0}.
\end{eqnarray}
The equality $m=M$ postulated by EP relates  $\varepsilon_0$ and $l_0$ as follows:
\begin{equation}
\varepsilon_0=\hbar \sqrt{\frac{g}{2l_0}},
\end{equation}
or, using the gravitational time scale Eq.(\ref{tau0}):
\begin{equation}
\tau_0= \sqrt{\frac{2l_0}{g}}.
\end{equation}
One can easily recognize in the above expression  a classical time of fall from the height $\l_0$ in the Earth's
gravitational field.

\begin{figure}[h!]
 \centering
\includegraphics[width=100mm]{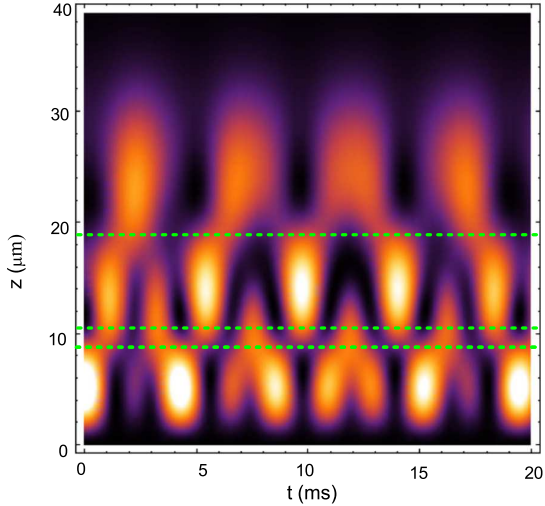}
\caption
{Color. The  probability density  of $\bar{H}$ in a superposition of the first,  second and third gravitational states,
as a function of the height $z$ above a mirror (vertical-axis) and the time $t$ (horizontal-axis).
Dark shade: low probability density. Light shade: high probability density.
The dashed lines indicate the positions of the nodes in the wave-functions of the second and third state.}
\label{Fig3zt}
\end{figure}

 Thus a measurement of the temporal-spatial probability density dependence of $\bar{H}$ in a superposition of the two
 lowest gravitation states would provide a full information on the gravitational properties of antimatter.
The superposition of three (and more) gravitational states could be useful to search for additional (to gravity)
interactions with a spatial scale of the order of $l_0$. For such a purpose, it is useful to study the probability
density at zeros of each   Airy function in the state superposition. In particular, the positions corresponding to
zeros of second and third gravitational states are  the following:
 \begin{eqnarray}
 z_1^{(2)}&=&(\lambda_2-\lambda_1)l_0=10.27 \mu m,\\ \label{z12}
 z_1^{(3)}&=&(\lambda_3-\lambda_2)l_0=8.41 \mu m, \\ \label{z13}
 z_2^{(3)}&=&z_1^{(2)}+z_1^{(3)}=18.68 \mu m, \label{z23}
 \end{eqnarray}
  The three state $(ijk)$ superposition probability density at  position of zero $z_i^k$  is equal to the two state
  superposition $(jk)$ probability:
  \begin{equation}
  |\Phi_{(ijk)}(z_n^{(k)},t)|^2=|\Phi_{(ij)}(z_n^{(k)},t)|^2.
  \end{equation}
  This means that  the three state probability density exhibits harmonic time-dependent oscillation with a frequency
  $\omega_{ij}$ at height of zero $z_n^k$. Let us mention that the time dependence of $|\Phi_{ijk}(z,t)|^2$ in any
  position $z$, except for the mentioned zeros is not harmonic; it is given by a superposition of three  cosine
  functions with different frequencies, analogous to Eq(\ref{3st}). This property allows us to extract the zeros
  positions using the probability density. One can see that the knowledge of zeros (Eq.(\ref{z12}-\ref{z23})) is
  analogous to the knowledge of the transition frequencies. A measurement of  one zero position allows us to extract
  the spatial scale $l_0$, a measurement of positions of two or more zeros allows us to constrain hypothetical nonlinear additions to the gravitational potential. We show the probability density  as a
  function of  height $z$ above the mirror (y-axis) and the time $t$ (x-axis) in a superposition of first, second and
  third gravitational state in Fig.\ref{Fig3zt}.

\section{Feasibility of antihydrogen gravitational states experiment}\label{Feas}
 In this  section we study  the feasibility of an experiment on the gravitationally bound quantum states of anti-hydrogen
 atoms. For such an estimation, we compare it with the already performed gravitational states experiments using
 ultra-cold neutrons (UCN) \cite{nesv02,nesv03,nesv05}. The UCN gravitational experiments can be used as a benchmark
 for such a comparison because 1) the neutron mass is nearly equal to the anti-hydrogen mass, 2) any modification to
 the $\bar{H}$ quantum state energies and wave-functions following from the precise shape of the Casimir-Polder potential
 are small compared to those of the quantum bouncer; UCN in the Earth's gravitational field above a perfect mirror is
 well described by the quantum bouncer model, 3) our estimation of $\bar{H}$ lifetimes in the quantum states of $0.1$ s
 are compatible to or even longer than the time of UCN passage through the mirror-absorber installation, also 4) UCN
 velocities are comparable to  velocities of ultra-cold $\bar{H}$ atoms produced in traps \cite{cesa05, gabr10}.  We
 will discuss here mainly the statistical limitations arising from an estimate of the
spectra from sources of $\bar{H}$ atoms that are projected in the near
future.

In the simplest configuration, the experimental method for observation of the neutron  gravitational states consisted
in  measuring the UCN flux through a slit between the horizontal mirror  and the flat absorber (scatterer) placed above
it at a variable height as a function of the slit height (the integral measuring method), or analyzing the spatial UCN
density distribution behind the horizontal bottom mirror exit (the differential measuring method) using
position-sensitive neutron detectors. The slit height can be changed and precisely measured. The absorber acts
selectively on the gravitational states, namely the states with a spatial size $H_n=\lambda_n^0 l_0$ smaller than the
absorber height $H$ are weakly affected, while the states with $H_n>H$ are intensively absorbed \cite{NVP,Adh,West}. A
detailed description of the experimental method, the experimental setup, the results of various applications of this
phenomenon could be found, for instance in  \cite{nesv02,nesv03,nesv05,UFN}.

Leaving aside numerous methodical difficulties in the experiments of this kind (as they have been already overcome in
the neutron experiments) and a real challenge to get high phase-space densities of trapped anti-hydrogen atoms (they
are aimed at anyway in the existing anti-hydrogen projects \cite{ATRAP,ALPHA}), let us compare relevant phase-space densities in the
two problems, keeping in mind that it is the principle parameter, which defines population of quantum states in accordance with the Liouville theorem. If the phase-space densities of anti-hydrogen atoms would be equal to those of UCN, we would just propose
to use an existing UCN gravitational spectrometer \cite{nesv00,kreuz} for anti-hydrogen experiments with minor
modifications.

UCN is an extremely narrow initial fraction in a much broader, and hotter neutron velocity distribution. Maximum UCN
fluxes available today for experiments in a flow-though mode are equal to $4$ $10^3$ $UCN/cm^2/s$; such UCN populate
uniformly the phase-space up to  the so-called critical velocity of about $6$ $m/s$ (UCN with smaller velocity are totally reflected from surface under any incidence angle; thus they could be stored in closed traps and transported using UCN guides). If one uses pulsed mode with a duty cycle,
say $10^{-3}$, the average flux would drop to $4$ UCN$/cm^2/s$. The pulse method provides more precise measurements, it
is used in current experiments with the GRANIT spectrometer, and it will be used in gravitational interference measurements
analogous to those performed with the centrifugal quantum states of neutrons \cite{Nature10,CentrNJP}. Taking into account
the  phase-space volume available for UCN in the gravitationally bound quantum states in the GRANIT spectrometer
\cite{kreuz}, we estimate the total count rate of about $10^2$ events/day when the relative accuracy for the
gravitational mass is  $10^{-3}$ (we note that the accuracy in the mentioned experiment is defined by a
width of a quantum transition, and a few events might be sufficient to observe the corresponding resonance).

The average flux of $\bar{H}$ atoms projected by AEGIS collaboration \cite{Aegis1, Aegis2} is a few atoms per second; let's take it equal $3$ $\bar{H}/s$ to have it defined. The cloud length is $\sim 8$ mm, its radius is $\sim1.5$ mm, thus the cloud volume is $\sim 5\times10^{-2}$ $cm^3$. For comparison, estimate of the average UCN flux, which would be emitted from a small UCN source with a volume of $\sim 5\times10^{-2}$ $cm^3$, the maximum UCN density available of $30$ $UCN/cm^3$, with a duty cycle of $\sim 10^{-3}$ gives $0.15$ $UCN/s$. It is $20$ times lower than the  $\bar{H}$ flux estimated above. One should not forget, however, that the projected $\bar{H}$ temperature is $\sim 100$ $m$K, i.e. $100$ times larger than the effective UCN temperature. Thus we lose a factor of $\sqrt{100}=10$ because of larger spread of $\bar{H}$ vertical velocities. No geometrical factors are taken into account here as well as no constraints following from final solid angles allowed, final sizes of $\bar{H}$ detectors, mirrors etc. However, their account would decrease our estimation by only a few times provided proper experiment design (note that equal acceleration of all anti-hydrogen atoms would not decrease their phase-space density). Thus we could provide statistical power of $\bar{H}$ experiment compatible to that with UCN. As the projected temperature of anti-hydrogen atoms in another proposal \cite{Yam} is significantly lower ($\sim 1$ $m$K that is just equal to the effective UCN temperature), the phase-space density of anti-hydrogen atoms could be even higher. Another significant advantage of a lower temperature consists in a more compact setup design. Note that a gravitational spectrometer analogous to \cite{nesv00,kreuz} selects just a very small fraction of UCN( $\bar{H}$) available (those with extremely small vertical velocity components) thus the count rate of "useful" events is extremely low in both cases.

Thus, we conclude that measurements of the gravitationally bound quantum states of anti-hydrogen atoms look realistic
if they would profit from methodical developments available in neutron experiments plus high phase-space densities of
anti-hydrogen atoms aimed at in future. Based on extensive analysis of the mentioned neutron experiments, we could
conclude that measurements of the gravitational mass of anti-hydrogen atoms with an accuracy of at least $10^{-3}$ is
realistic, provided that projected high $\bar{H}$ phase-space density is achieved.

\section{Conclusions}
We argue on existence of long-living quasi-stationary states of $\bar{H}$  above a material surface in the
gravitational field of Earth. A typical lifetime of such states above an ideally conducting plane surface is
$\tau\simeq 0.1$ s. Quasi-stationary character of such states is  due to the quantum reflection of ultra-cold
(anti)atoms from the Casimir-Polder (anti)atom-surface potential. The relatively long life-time is due to the smallness
of the ratio of the characteristic spatial antiatom-surface interaction scale $l_{CP}$ and the spatial gravitational
scale $l_0$. We show that the spectrum of $\bar{H}$ decaying gravitational levels is quasi-discrete even for the highly excited  states as long as their quantum number $n\ll 30 000$. We argue that
low lying gravitational states provide an interesting  tool for studying the  gravitational properties of antimatter,
in particular for testing the equivalence between the gravitational and inertial masses of $\bar{H}$.
We show that,
by counting the number of $\bar{H}$ annihilation events on the surface,
both the transition frequencies between the gravitational  energy levels, as well as the spatial density distribution
of gravitational states superposition can be measured. An important observation in this context is
that a modification of the above mentioned properties of gravitational states due to the interaction with a surface
disappears in the first order of the small ratio $l_{CP}/l_0$.  Finally, we show that actual measurements  of quantum
properties of $\bar{H}$ atoms, levitating above a material mirror in gravitational states are feasible,  provided
that the projected high  phase-space density $\bar{H}$ is achieved.

\section {Acknowledgments}
We  would like to acknowledge the support from the
Swedish  Research Council, from the  Wenner-Gren
Foundations and  the Royal Swedish Academy of Sciences.

\section{Appendix}
Here we derive an expression for the scalar product of two complex energy gravitational states eigenfunctions:
\begin{equation}
\alpha_{ij}=\frac{1}{N_iN_j}\int_0^\infty\mathop{\rm Ai^*}(z/l_0-\lambda_{j})\mathop{\rm Ai}(z/l_0-\lambda_i)dz
\end{equation}
with
\begin{equation}\label{normdef}
N_i^2=\int_0^\infty \mathop{\rm Ai^2}(z/l_0-\lambda_{j})dz.
\end{equation}
We start with the equations for the eigenfunction $\mathop{\rm Ai}(z/l_0-\lambda_i)$ and the complex eigenvalues
$\lambda_i=\lambda_{i}^0+a_{CP}/l_0$:

\begin{equation} \label{Lambda1Eq}
-\mathop{\rm Ai''}(z/l_0-\lambda_i)+z\mathop{\rm Ai}(z/l_0-\lambda_i)=\lambda_i \mathop{\rm Ai}(z/l_0-\lambda_i).
\end{equation}
The equations for the complex conjugated eigenfunction and the eigenvalue are:
\begin{equation}\label{Lambda2Eq}
-\mathop{\rm Ai^{*''}}(z/l_0-\lambda_j)+z\mathop{\rm Ai^*}(z/l_0-\lambda_j)=\lambda_{j}^{*} \mathop{\rm
Ai^*}(z/l_0-\lambda_j).
\end{equation}
We multiply both sides of equation Eq.(\ref{Lambda1Eq}) by $\mathop{\rm Ai^*}(z/l_0-\lambda_{j})$ and integrate them
over $z$. Then we multiply both sides of Eq.(\ref{Lambda2Eq}) by $\mathop{\rm Ai}(z/l_0-\lambda_{i})$ and integrate
them over $z$. After substraction of the results of these operations we get:
\begin{equation}
\mathop{\rm Ai^{*'}}(-\lambda_{j})\mathop{\rm Ai}(-\lambda_{i})-\mathop{\rm Ai^{*}}(-\lambda_{j})\mathop{\rm
Ai'}(-\lambda_{i})=(\lambda_j^*-\lambda_i)\int_0^\infty\mathop{\rm Ai^*}(z/l_0-\lambda_{j})
\mathop{\rm Ai}(z/l_0-\lambda_i)dz.
\end{equation}
 To get the above result, we integrated by parts the integrals with second derivatives and took into account that Airy
 functions vanish at infinity.
 Now we take into account the equality $\mathop{\rm Ai}(-\lambda_i^0)=0$ and  smallness of the ratio $a_{CP}/l_0$, to
 get the following expressions, exact up to the second order in  $a_{CP}/l_0$:
 \begin{eqnarray}
 \mathop{\rm Ai}(-\lambda_i)=-\frac{a_{CP}}{l_0}\mathop{\rm Ai'}(-\lambda_i^0),\label{Lambda1Bc}\\
 \mathop{\rm Ai^*}(-\lambda_j)=-\frac{a_{CP}^*}{l_0}\mathop{\rm Ai^{*'}}(-\lambda_j^0).\label{Lambda2Bc}
 \end{eqnarray}
Up to the second order in $a_{CP}/l_0$ we get:
\begin{equation}
\mathop{\rm Ai^{*'}}(-\lambda_{j})\mathop{\rm Ai}(-\lambda_{i})-\mathop{\rm Ai^{*}}(-\lambda_{j})\mathop{\rm
Ai'}(-\lambda_{i})=i\frac{b}{2l_0}\mathop{\rm Ai^{*'}}(-\lambda_{j}^0)
\mathop{\rm Ai'}(-\lambda_{i}^0).
\end{equation}
The above result should be combined with the known expression for the normalization coefficient:
\begin{equation}\label{normex}
N_i^2=\mathop{\rm Ai'^2}(-\lambda_{i})+\lambda_i\mathop{\rm Ai^2}(-\lambda_{i}),
\end{equation}
which,  up to the second order in  $a_{CP}/l_0$ turns to be:
\begin{equation}\label{norm}
N_i=\mathop{\rm Ai'}(-\lambda_{i}^0)(1+\lambda_i^0\frac{a_{CP}^2}{2l_0^2}).
\end{equation}
Keeping first order terms, we finally  get:
\begin{equation}\label{crossT}
\alpha_{i\neq j}=i\frac{b/(2l_0)}{\lambda_j^0-\lambda_i^0+i b/(2l_0)}.
\end{equation}

\bibliographystyle{unsrt}
\bibliography{hbarclock}

\end{document}